# Code C# for chaos analysis of relativistic many-body systems

I.V. Grossu[a,*], C. Besliu[a], Al. Jipa[a], C.C. Bordeianu[a], D. Felea[b], E. Stan[b], T. Esanu[a]

[a] *University of Bucharest, Faculty of Physics, Bucharest-Magurele, P.O. Box MG 11, 077125, Romania*
[b] *Institute of Space Sciences, Laboratory of Space Research, Bucharest-Magurele, P.O. Box MG 23, 077125, Romania*

### ABSTRACT

This work presents a new Microsoft Visual C# .NET code library, conceived as a general object oriented solution for chaos analysis of three-dimensional, relativistic many-body systems. In this context, we implemented the Lyapunov exponent and the "fragmentation level" (defined using the graph theory and the Shannon entropy). Inspired by existing studies on billiard nuclear models and clusters of galaxies, we tried to apply the virial theorem for a simplified many-body system composed by nucleons. A possible application of the "virial coefficient" to the stability analysis of chaotic systems is also discussed.

## Program summary

*Manuscript title: Code C# for chaos analysis of relativistic many-body systems*
*Authors: I.V.Grossu, C. Besliu, Al. Jipa, C.C. Bordeianu, D. Felea, E. Stan ,T. Esanu*
*Program title: Chaos Many-Body Engine v01*
*Licensing provisions:*
*Programming language: Visual C# .NET 2005*
*Computer(s) for which the program has been designed: PC*
*Operating system(s) for which the program has been designed: .Net Framework 2.0 running on MS Windows*
*RAM required to execute with typical data: 80 Megabytes*
*Has the code been vectorised or parallelized?: each many-body system is simulated on a separate execution thread*
*Number of processors used: one per simulated many-body system*
*Supplementary material: Windows forms application for testing the engine.*
*PACS:*
*CPC Library Classification:*
*External routines/libraries used: .Net Framework 2.0 Library*
*Nature of problem: Chaos analysis of three-dimensional relativistic many-body systems.*
*Solution method: Second order Runge-Kutta algorithm for simulating relativistic many-body systems. Object oriented solution, easy to reuse, extend and customize, in any development environment which accepts .Net assemblies or COM components. Implementation of: Lyapunov exponent, "fragmentation level", "average system radius", "virial coefficient", and energy conservation precision test.*
*Restrictions:*
*Additional comments: Easy copy/paste based deployment method.*
*Running time: quadratic complexity*

**1. Introduction**

Based on a second order Runge-Kutta algorithm [1], we developed a code library for the simulation of relativistic many-body systems [2]. Our attention was mainly focused on both creating a general object oriented solution [1,3,4], easy to customize and extend through the inheritance and polymorphism mechanisms, and developing a set of tools for analyzing the chaotic behavior of many-body systems.

The main reasons for choosing Visual C# .Net [4] are related to the high advantages provided by an object oriented language, together with the Microsoft .NET technology: code reusability, managed runtime environment, rapid development of windows forms applications, XCOPY deployment strategy [4], benefit from all functionalities included in the .Net framework library, etc.

Over the last two decades an increasing number of papers have treated the deterministic chaotic behavior of Fermi nuclear systems [5-7]. In this context, we used the application for the analysis of a simplified many-body system composed by nucleons.

**2. Program description**

The "Chaos Many-Body Engine" library is designed as a general numerical solution for the simulation of three-dimensional relativistic many-body systems. It is based on two layers: the application logic layer (*Engine.dll*), and the data access layer (*Data.dll*). Following the belief that the user should not treat the program as a "black-box", an important attention was paid to the application extensibility.

The *Math.dll* library contains various functionalities of general interest. The *Vector* class is used for implementing three-dimensional vectors. Its *X, Y, and Z* properties represent the three Cartesian components. The +, -, *, / operators were overloaded in order to implement the basic vector operations. The *Relativity* class contains a set of static methods for the most common relativistic relations (obtain the movement mass from rest mass and velocity; obtain the velocity from momentum and rest mass etc.). The *Graphic* class stores a set of *(x, f(x))* pairs, and is mainly used for graphical representations.

The *Engine.dll* library offers a framework for many-body simulations. Starting from the relativistic generalization of the second Newton's law, we considered the following system of equations:

$$\begin{cases} \dfrac{d\vec{p}_i}{dt} = \sum_{i <> j} \vec{F}_{ij} = \vec{F}_i \\ m_i \dfrac{d\vec{r}_i}{dt} = \vec{p}_i \qquad (1) \\ m_i = \dfrac{m_{0i}}{\sqrt{1 - \left(\dfrac{v_i}{c}\right)^2}} \end{cases}$$

where, $p_i$ is the momentum, $r_i$ the position, $v_i$ the velocity, $m_{0i}$ the rest mass, $m_i$ the movement mass of the constituent $i$, $t$ is the time, $F_{ij}$ the bi-particle force, and $c$ is the velocity of light in vacuum. The collisions could be implemented by employing an appropriate repulsive term in the force expression.

For assuring the application flexibility, the user is allowed to programmatically define all the specific properties of the desired many-body system: the number of constituents, the properties of each particle, the initial positions and velocities, the expression of the bi-particle interaction, and the system type (classical or relativistic). The application was mainly conceived for working in the center of momentum reference frame. Following this purpose, the programmer should set the initial conditions accordingly.

The equations (1) are solved using a second order Runge-Kutta algorithm [1]. We calculated first the resultant force $F_i(t_n)$ acting on each particle. We implemented also a "fast" version of the algorithm, in which $F_i(t_n)$ is approximated with the previous calculated force $F_i(t_{n-1/2})$. Based on this information, one can obtain the positions $r_{in+1/2}$, corresponding to the midpoint of the integration interval $h$ (Euler method). The resultant force $F_i(t_{n+1/2})$, considered for this intermediary status, was used for computing the new momentum $p_{in+1}=p_{in}+h.F_i(t_{n+1/2})$. Based on the average velocity value $v_{iavg}=(v_{in}+v_{in+1})/2$, we obtained the new position $r_{in+1}=r_{in}+h.v_{iavg}$.

The energy conservation is not explicitly built in the algorithm. The precision could be thus checked by plotting [1]

$$-log_{10}\left|\frac{E(t)-E(t=0)}{E(t=0)}\right| \quad (2)$$

for the entire simulation time. For performing the constancy of energy test, the user must implement also the expression of the bi-particle potential.

The *Particle* class abstracts a material point. It encapsulates some basic, scalar and vector, properties (rest mass, movement mass, electric charge, position, velocity, momentum, and force). The *Clone* method returns a copy of the current instance. More properties could be added by inheriting the class.

The *NBody* class encapsulates a set of interacting material points (generic list of *Particle* objects). For the bi-particle interaction expression we implemented two type-safe function pointers (delegates [4]): *Force* and *Potential*. The programmer could thus provide any relation, but is also enforced to conform to the required function prototype.

The *Next* method implements the, previously described, Runge-Kutta algorithm for solving the equations (1). It has two overloaded versions, one for obtaining the next status of a single particle, and another for "moving" the entire system into its next status. For assuring the flexibility, both versions are defined as virtual functions. The programmer has thus the possibility to modify the default behavior of the engine.

An important attention was paid to the code optimization. Taking also into account the Newton's third law, the algorithm complexity is given by:

$$C_{rk2}=n(n-1)$$
$$C_{rk2\ fast}=\frac{n(n+1)}{2} \quad (3)$$

The *World* class implements a collection of independent many-body systems (generic list of *NBody* objects). Most commonly, we used it for studying the "butterfly effect" [8], by simulating two identical systems with slightly different initial conditions.

The *Start* method contains the main loop of the simulation process. On each iteration, the *Next* method is called for all *NBody* instances stored in the *mSystems* generic list variable. The external applications are notified on the calculation progress through the *Compute* event, and on the simulation end through the *SimulationEnd* event. The *Start* function has also an overloaded version which takes no parameters. It is used, together with the *SetParams* method, for initiating the simulation parallel regime (one execution thread per system). The *Stop* method triggers the simulation end.

The abstract class *SimulationBase* provides the engine programmatic interface for developers, and is conceived as a "contract" which enforces a proper implementation of the specific many-body system of interest. One can set the initial conditions by overriding the virtual method *SetInitialStatus*, while the abstract function *Force* imposes the implementation of the bi-particle interaction. For the constancy of energy test (2), the programmer should provide also the potential expression (overriding the *Potential* function). There is no restriction for deriving the force from potential by using an appropriate numerical method. However, explicitly providing both expressions is recommended for performance reasons.

The *Start* method calls the *SetInitialStatus* method, assigns the force and potential functions to each *NBody* object, and starts the simulation by calling the *Start* method of the *mWorld* variable (instance of the *World* class). The *StartParallel* method instantiates one *World* object for each *NBody* system, and starts each simulation on a separate execution thread. The benefits become visible on a multi-core processor. The *Stop* method triggers the simulation end. The *Started* property returns true when at least one simulation is still executing. The *OnCompute* and *OnSimulationEnd* virtual functions are used as handlers for the, previously mentioned, *NBody.Compute*, respectively, *NBody.SimulationEnd* events.

The data access library (Data.dll) can be used independently, and provides basic processing functionalities and programmatic access to the application output. Each simulation is stored into a set of comma separated values files (text format recognized also by Microsoft Excel). For each *NBody* object there are two files: the header (containing general information: particle index, rest mass, electric charge, initial position, and initial momentum), and the data file (which contains information on the system evolution: particle index, position, momentum, and time). The file names conform to the following naming rule: {FileName}.{Type}.{SystemIndex}.csv, where "Type" could be "dat" or "hdr", and "SystemIndex" represents the index of the *NBody* object. The information is stored in the mentioned files by the *SaveHeader* and *Save* methods of the *NBody* class. In order to avoid huge file generation, the user has the possibility to specify the frequency of the *Save* function call.

The *BaseSystemData* class is used for loading the simulation output into a dataset object [4]. The programmatic access to the information of interest is assured by a specific set of methods and properties: *Count, Mass, Charge, InitialPosition, InitialMomentum, Time, Position, and Momentum*. The *SystemData* class, defined in the *Engine.dll* library, extends *BaseSystemData*. Its *GetInstance* method returns an *NBody* object which represents a "snapshot" of the system, considered at a given moment of time.

### 3. Chaos analysis of many-body systems

A qualitative measure of whether a dynamical system is chaotic or not is given by the Lyapunov exponents. For analyzing the instability on perturbations, the program facilitates the simulation of two identical systems with slightly different initial conditions. We considered the multi-dimensional Lyapunov Exponent, defined as [9,10]:

$$L \stackrel{\text{def}}{=} \lim_{t \to \infty} \frac{1}{t} ln \frac{d(t)}{d(0)} = \lim_{t \to \infty} L(t) \quad (4)$$

where *d(t)* represents the phases space distance between the two systems. The Lyapunov Exponent offers a qualitative image of the system behavior (for *L>0* the neighboring orbits in phase space diverge and are chaotic; for *L=0* the orbits remain marginally stable; for *L<0* the orbits converge). The *LyapunovExponent* function was implemented as a static method of the *SystemData* class. It expects two *SystemData* objects as parameters, and returns a *Graphic* object containing the dependence on time of *L(t)*.

From a computational perspective, one practical idea could be to apply the graph theory [11] in the study of Many-Body system. Thus, at a given moment of time, one can consider the graph *G(C,I)*, where *C* is the set of particles, and *I* is the set of particle pairs for which the distance is lower than the interaction radius. In this context, it is possible to define the notion of cluster as a maximal connected sub-graph of *G* [12]. Based on the Shannon entropy [10,13], one can define the "fragmentation level" of one Many-Body system as [12]:

$$F(t) \stackrel{\text{def}}{=} -\sum_{i=1}^{N_C(t)} f_i(t) \ln\bigl(f_i(t)\bigr) \; with \; f_i(t) = \frac{n_i(t)}{n} \quad (5)$$

where *t* is the time, $N_C$ the number of clusters, $n_i$ the constituents number of the cluster *i*, and *n* is the total number of particles. *F* offers a quantitative measure of the fragmentation degree of one many-body system. It relates the existing number of clusters by their dimensions. *F* is an additive, monotonic function, which equals zero when the system acts as a single cluster, and reaches the maximum value when the system is totally decomposed into its elementary constituents.

The *GetSubCluster* method of the *NBody* class is used for obtaining the maximal connected sub-graph of *G*, starting from an arbitrary particle, specified as parameter. We implemented the following algorithm:
1. The starting particle is added to the cluster, and the *bChanged* boolean variable is initialized with true;
2. If there was no modification (*bChanged=false*), the algorithm ends. The *cluster* variable contains the result.
3. *bChanged = false*
4. For each particle *p*, which is not already included in any cluster, if it exists at least one particle *q* inside the cluster, so that *distance(p,q)<interaction radius*, then *p* is added to the cluster, and *bChanged* becomes true
5. go to 2

Based on the *GetSubCluster* function, the *GetClusters* method returns the list of all maximal connected sub-graphs of *G*. The *FragmentationLevel* method uses this information for computing the, previously defined, fragmentation level *F* (5).

Fritz Zwicky was the first to use the virial theorem for deducing the existence of dark matter [14]. The main advantage of this theorem is that it does not depend on the notion of temperature and can be applied even for systems which are not in thermal equilibrium. For the relativistic case, one can define the "virial coefficient" as:

$$Vis \stackrel{\text{def}}{=} \lim_{t \to \infty} \frac{\langle \sum_k p_k \cdot v_k \rangle}{\langle \sum_k F_k \cdot r_k \rangle} \quad (6)$$

where the <> operator denotes an average over the time, $p_k$ is the momentum, $v_k$ the velocity, $r_k$ the position, and $F_k$ the force acting on the constituent *k*. According to the virial theorem, the *Vis* coefficient is unitary for a bound system and greater than unity when the system expands. Another advantage is related to the fact that the virial coefficient is less depending on the algorithm precision.

As the information on force is not stored in the output files, the two terms *Sum(p_k,v_k)*, and *Sum(F_k r_k)* are calculated during the simulation, in the *Next* method of the *NBody* system. The corresponding values are accessible through the *PkVk* and *FkRk* properties. For calculating the average values, two variables, *mAvgFkRk* and *mAvgPkVk*, were defined in the *World* class. The two variables are actualized, on each iteration, in the simulation main loop. The value of the virial coefficient is thus accessible through the read only property *World.Vis*, only after the simulation ends.

An intuitive image on the evolution type (static, expansion, collapse), is given also by the dependence on time of the system's average radius, defined as [12]:

$$R(t) = \sum_{i<j} \frac{r_{ij}(t)}{n(n-1)} \quad (7)$$

where $r_{ij}$ is the bi-particle distance, and *n* is the number of constituents.

The static method *AverageDistance* of the *Vector* class is used for computing the average radius for a set of points, specified as an array of *Vector* variables. The *BaseSystemData.AvgDistance* method returns a *Graphic* object containing the dependence on time of *R(t)*.

One can notice that the phases-space distance between two many-body systems cannot overcome a maximal value if the systems are bound. In these conditions, the limit (4) will equal zero, even for a system with chaotic behavior [6]. By renouncing to the *1/t* factor, the expression could still tend to zero for a chaotic system which collapses. Trying to get proper information, not affected by the evolution type (expansion, bound state, collapse), one idea could be to introduce a dependence on the virial coefficient in the definition of the Lyapunov exponent. In this context, it is important to emphasis on the intuitive nature of relation (8). More studies and tests must be provided in order to confirm its validity.

$$L_{Vis} \stackrel{\text{def}}{=} \lim_{t\to\infty} \frac{1}{t^{(Vis-1)}} ln \frac{d(t)}{d(0)} \quad (8)$$

**4. Application to a nuclear toy-model**

Inspired by existing studies on billiard nuclear models [5-7], we used the previously described program for chaos analysis of a simplified many-body system composed by nucleons [12]. A rudimentary windows forms application was developed following this purpose.

We employed a finite depth Yukawa potential well, together with a coulombian term:

$$V(r_{ij}) = \begin{cases} -V_0, & r_{ij} < 0.567a \\ -V_0 \frac{e^{-\frac{r_{ij}}{a}}}{\frac{r_{ij}}{a}} + \frac{q_i q_j}{4\pi\epsilon r_{ij}}, & r_{ij} \geq 0.567a \end{cases} \quad (9)$$

where $V_0=35MeV$ is the depth and $a=2Fm$ is the radius of the potential well, $r_{ij}$ represents the distance between the two bodies, and *q* is the electric charge.

The particles are initially placed in the vertices of a regular centered dodecahedron with radius $r=1.5Fm<r_0 A^{1/3}=1.2*21^{1/3}Fm=3.3Fm$ (compressed nucleus). For simplicity, we considered a homogenous mix of protons and neutrons, with radial initial velocities (explosion model).

By representing the fragmentation level (5) and the virial coefficient (6), measured at the maximum simulation time ($t_{max}=200Fm/c$), as functions of initial velocity (Fig.1), one can notice the existence of three regions:

- A first region ($v_{ini} \approx 0$ to $0.4c$), corresponding to a bound system, where the virial coefficient is practically constant and equals one. The fragmentation level is close to zero, as only a few particles could occasionally escape from the system. An intuitive image of the oscillations regime is given by the average radius dependence on time (7), considered for $v_{ini}=0$ (Fig.2:Right);
- An intermediary region ($v_{ini} \approx 0.4c$ to $0.55c$), corresponding to a quasi-linear increasing of the fragmentation level with the initial velocity (presence of clusters). One can intuitively connect this behavior with the nuclear fragmentation mechanism [15].
- A region ($v_{ini} \geq 0.55c$) where the virial coefficient is significantly higher than unity (expansion) and increases fast with the initial velocity. The fragmentation level is maximal as the system is decomposed into its elementary constituents.

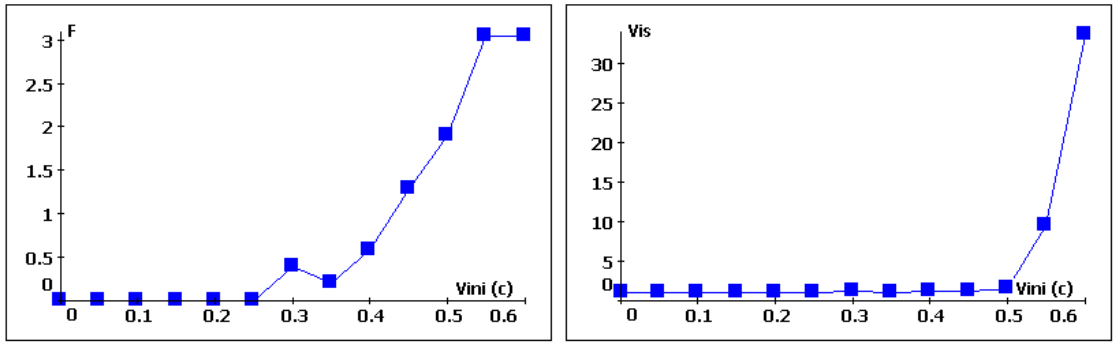

**Fig.1.** (Left) The fragmentation level (5) as a function of initial velocity. (Right) The virial coefficient (6) as a function of initial velocity. (The lines which connect points are only for eye orientation.)

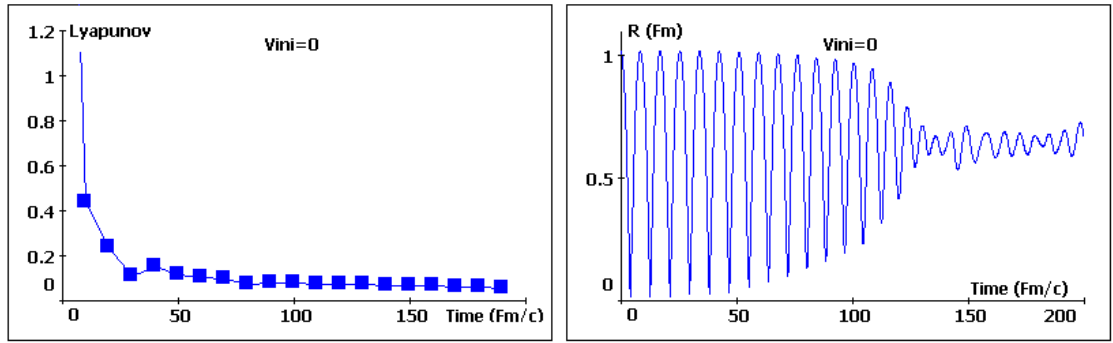

**Fig.2.** (Left) The Lyapunov Exponent (4) as a function of time, for $v_{ini}=0$. (Right) The average system radius (7) as a function of time, for $v_{ini}=0$. (The lines which connect points are only for eye orientation.)

**5.Conclusion**

We developed a managed [4], object oriented many-body simulation engine, easy to reuse extend and customize through the inheritance and polymorphism mechanisms. It was conceived as a set of libraries which can be used on a wide scale, in any development environment which accepts .Net assemblies or COM components. The access to the application output is simplified by using comma separated values files. The data layer module can be used independently, and provides processing functionalities and programmatic access to the engine output.

For studying the deterministic chaotic behavior of many-body systems, we implemented the Lyapunov exponent and, based on the graph theory and the Shannon entropy, we defined the "fragmentation level" (5). A possible application of the virial theorem in the stability analysis of non-linear dynamical systems is also discussed (8).

In the last years, an important attention was accorded to the chaos analysis of Fermi nuclear systems. In this context, we discuss the relativistic virial coefficient (6), as an additional tool which does not depend on the notion of temperature, and remains valid even for systems far from thermal equilibrium.

**Appendix A. Chaos-Many Body Engine example of use**

For implementing the, previously described, nuclear system, we created a test windows forms C# project (*ManyBody.csproj*), we added references to the engine libraries (*Math.dll, Data.dll, Engine.dll*), and we inherited the *SimulationBase* class. There is no restriction for using the engine in any other environment which accepts .Net assemblies or COM components (e.g. a web based application [16]).

The *GetSystem* function is used for setting the specific properties of the considered many-body system. For analyzing the "butterfly effect", in the *SetInitialStatus* method we added two identical systems with slightly different dodecahedron radius. We provided the expression for both the bi-particle force and the bi-particle potential. The output of the program, for *Radius=1.5Fm and InitialVelocity=0*, is ploted in (Fig.2).

```csharp
class SimulationRelativisticExample : SimulationBase
{
    double coulomb = 1.44;
    double mRadius = 0;
    double mRadialVelocity = 0;
    double mp = 938.26; //proton mass in MeV
    double mn = 939.55; //neutron mass in MeV

    public double Radius
    {
        get
        {
            return mRadius;
        }
        set
        {
            mRadius = value;
        }
    }

    public double InitialVelocity
```

```csharp
        {
            get
            {
                return mRadialVelocity;
            }
            set
            {
                mRadialVelocity = value;
            }
        }

        protected override Vector Force(Particle p, Particle q)
        {
            Vector v = (p.Position - q.Position);
            double r = Math.Abs(v.Module);
            double a = 2;
            double ra = r / a;
            double V0 = 35;
            Vector F = new Vector();
            double delta = 0.567 * a;
            //avoiding huges values for short distances
            if (r > delta)
            {
                //Yukawa
                F = F + ((-V0 / a) * Math.Exp(-ra) * (ra + 1) / (ra * ra)) * v.Versor;
                //Coulombian term
                F = F + (coulomb * p.Charge * q.Charge / (r * r)) * v.Versor;
            }
            return F;
        }

        protected override double Potential(Particle p, Particle q)
        {
            Vector v = (p.Position - q.Position);
            double r = Math.Abs(v.Module);
            double a = 2;
            double ra = r / a;
            double V0 = 35;
            double V = 0;
            double delta = 0.567 * a;
            V = -V0;
            //avoiding huges values for short distances
            if (r > delta)
            {
                //Yukawa
                V = -V0 * Math.Exp(-ra) / ra;
                //Coulombian term
                V += coulomb * p.Charge * q.Charge / r;
            }
            else
            {
                V = -V0 * Math.Exp(-delta / a) / (delta / a) + coulomb * p.Charge * q.Charge / delta;
            }
            return V;
        }

        protected override void OnCompute(World obj)
        {
```

```csharp
            base.OnCompute(obj);
            //write your custom code for treating the Compute event
        }

        protected override void OnSimulationEnd(World obj)
        {
            base. OnSimulationEnd(obj);
            //write your custom code for treating the SimulationEnd event
        }

        protected override void SetInitialStatus()
        {
            base.SetInitialStatus();
            NBody s = null;
            double dr = 0.001;

            //obtains a new n-body system
            s = GetSystem(mRadius, mRadialVelocity);
            mWorld.Systems.Add(s);

            //obtains a new n-body system, with a slightly different initial dodecahedron radius
            s = GetSystem(mRadius + dr, mRadialVelocity);
            mWorld.Systems.Add(s);
        }

        private NBody GetSystem(double r, double velocity)
        {
            NBody sys = new NBody();
            Particle p = null;
            //compute the edge based on the dodecahedron's radius
            double edge = r * 0.714;

            //obtain the dodecahedron configuration
            //the initial radial velocity is also specified
            sys = Configuration.Dodecahedron(edge, velocity);
            //centered dodecahedron (for symmethry reasons)
            p = new Particle(0, 0, 0, 0);
            sys.Particles.Add(p);

            //set the properties for each particle (homogen mix of protons and neutrons)
            for (int i = 0; i < sys.Particles.Count; i++)
            {
                p = sys.Particles[i];
                //homogen mix
                if (Functions.Mod(i, 2) == 0)
                {
                    //proton
                    p.Mass = mp;
                    p.Charge = 1.0;
                    p.Velocity = velocity * p.Velocity.Versor;
                    p.ForeColor = System.Drawing.Color.Red;
                }
                else
                {
                    //neutron
                    p.Mass = mn;
                    p.Charge = 0.0;
                    p.Velocity = velocity * p.Velocity.Versor;
                    p.ForeColor = System.Drawing.Color.Blue;
```

```
            }
        }
        sys.Type = SystemType.Relativistic;
        return sys;
    }
}
```

**References**

[1] R.H. Landau, M.J. Paez, C.C. Bordeianu, Computational Physics: Problem Solving with Computers, Wiley-VCH-Verlag, Weinheim, 2007
[2] P. Ring and P. Schuck, The Nuclear Many Body Problem, Springer-Verlag, Berlin, 1980.
[3] Bjarne Stroustrup, The C++ programming language, AT&T, USA, 1997.
[4] Christian Nagel, Bill Evjen, Jay Glynn, Morgan Skinner, Karli Watson, Professional C# 2008, Wiley, Indianapolis, Indiana, 2008
[5] G.F. Burgio, F.M. Baldo, A. Rapisarda, P. Schuck, Phys. Rev. C 58 (1998) 2821–2830.
[6] D. Felea, PhD thesis, University of Bucharest, Physics Department, 2002.
[7] C.C. Bordeianu, D. Felea, C. Besliu, Al. Jipa, I.V. Grossu, Computer Physics Communications 179 (2008) 199–201.
[8] Zbyszek P. Karkuszewski, Christopher Jarzynski, and Wojciech H. Zurek, Quantum Chaotic Environments, the Butterfly Effect, and Decoherence, Physical Review Letters, volume 89, 17 (2002).
[9] M. Sandri, Numerical Calculation of Lyapunov Exponents, Mathematica J.6, 78-84, 1996.
[10] C.C. Bordeianu, D. Felea, C. Besliu, Al. Jipa, I.V. Grossu, Computer Physics Communications 178 (2008) 788-793.
[11] Alan Gibbons, Algorithmic Graph Theory, Cambridge University Press, 1985.
[12] I.V.Grossu, PhD thesis, University of Bucharest, Physics Department, 2010.
[13] Shannon, C. E. (1948), A mathematical theory of communication, Bell System Tech. J. 27, 379.
[14] Richard Panek, The Father of Dark Matter, Discover, January 2009, pp. 81-87.
[15] Al. Jipa, C.Besliu, D. Felea, et al., Romanian Reports in Physics, Volume 56, P. 577-601, No. 4, 2004.
[16] Andrei Ignat, Asp.NET MVC Tips and Tricks, CreateSpace, US, 2009